%% file: ms.tex
\pdfoutput=1 
\documentclass[conference]{IEEEtran}
\IEEEoverridecommandlockouts 

\makeatletter
\let\NAT@parse\undefined
\makeatother

\usepackage{amsmath,amssymb,amsfonts,amsthm,commath,esint}
\usepackage[numbers,sort&compress]{natbib}

\usepackage{mathtools}
\usepackage[nolist,nohyperlinks]{acronym}
\usepackage{color}
\usepackage{multirow}
\usepackage{enumitem}
\usepackage{physics}
\usepackage{optidef}
\usepackage{mleftright}
\usepackage{cuted}
\usepackage{algorithm2e}
\usepackage{microtype}

\SetKwInput{kwInit}{Initialization}

\ifCLASSOPTIONcompsoc
\usepackage[caption=false, font=normalsize, labelfont=sf, textfont=sf]{subfig}
\else
\usepackage[caption=false, font=footnotesize]{subfig}
\fi

\makeatletter
\def\blfootnote{\xdef\@thefnmark{}\@footnotetext}
\makeatother

\begin{acronym}
	\acro{RMS}{root mean square}
	\acro{IRS}{intelligent reflecting surfaces}
	\acro{SNR}{signal-to-noise ratio}
	\acro{nCA-MF}{non-coupling-aware matched filtering}
	\acro{WMF}{whitened matched filtering}
	\acro{CA-MF}{coupling-aware matched filtering}
	\acro{nCA-ZF}{non-coupling-aware zero forcing}
	\acro{CA-ZF}{coupling-aware zero forcing}
	\acro{SVD}{singular value decomposition}
	\acro{EMS}{electromagnetic surface}
	\acro{MF}{matched filtering}
	\acro{KKT}{Karush–Kuhn–Tucker}
	\acro{EIRP}{effective isotropic radiated power}
	\acro{AWGN}{additive white Gaussian noise}
	\acro{MMSE}{minimum mean square error}
	\acro{SINR}{signal-to-interference-plus-noise ratio}
	\acro{MIMO}{multiple-input multiple-output}
	\acro{LIS}{large intelligent surfaces}
	\acro{UE}{user equipment}
	\acrodefplural{UE}[UEs]{user equipments}
	\acro{LoS}{line-of-sight}
	\acro{ZF}{zero-forcing}
	\acro{WMMSE}{weighted minimum mean square error}
	\acro{CSI}{channel state information}
\end{acronym}

\renewcommand{\vec}[1]{\mathbf{\lowercase{#1}}}
\newcommand{\mat}[1]{\mathbf{\uppercase{#1}}}

\newcommand{\e}[1]{\mathrm{e}^{#1}}

\NewDocumentCommand{\evalat}{sO{\big}mm}{%
	\IfBooleanTF{#1}
	{\mleft. #3 \mright|_{#4}}
	{#3#2|_{#4}}%
}

\DeclarePairedDelimiterXPP\Aver[1]{\mathbb{E}}{[}{]}{}{
	
	#1
}

\begin{document}

\title{Multiuser MIMO with Large Intelligent Surfaces: Communication Model and Transmit Design\\
\thanks{This work has been supported by the Danish Council for Independent Research under grant DFF-701700271.}}

\author{\IEEEauthorblockN{Robin Jess Williams\IEEEauthorrefmark{1}, Pablo Ram\'irez-Espinosa\IEEEauthorrefmark{1}, Elisabeth de Carvalho\IEEEauthorrefmark{1} and Thomas L. Marzetta\IEEEauthorrefmark{2}}
\IEEEauthorblockA{\IEEEauthorrefmark{1}Department of Electronic Systems, Connectivity Section (CNT) Aalborg University, Denmark\\ \IEEEauthorrefmark{2}Tandon School of Engineering, New York University, Brooklyn, NY\\ Email: \IEEEauthorrefmark{1}\{rjw, pres, edc\}@es.aau.dk, \IEEEauthorrefmark{2}tom.marzetta@nyu.edu
}}

\maketitle

\input{abstract.tex}

\section{Introduction}\label{sec:introduction}
\input{introduction.tex}

\section{System model}\label{sec:systemModel}
\input{systemModel.tex}

\section{System analysis: coupling and received power}\label{sec:systemAnalysis}
\input{systemAnalysis.tex}

\section{MIMO communication model}\label{sec:MIMOmodel}
\input{MIMOcommunmodel.tex}

\section{Transmit design}\label{sec:transmitDesign}
\input{transmitDesign.tex}

\section{Numerical results}\label{sec:simulation}
\input{simulation.tex}

\section{Conclusions}\label{conclusion}
\input{conclusion.tex}

\bibliographystyle{IEEEtran}
\bibliography{IEEEabrv,IEEEexample}

\end{document}

%% file: abstract.tex
\begin{abstract}
This paper proposes a communication model for multiuser multiple-input multiple-output (MIMO) systems based on large intelligent surfaces (LIS), where the LIS is modeled as a collection of tightly packed antenna elements. The LIS system is first represented in a circuital way, obtaining expressions for the radiated and received powers, as well as for the coupling between the distinct elements. Then, this circuital model is used to characterize the channel in a line-of-sight propagation scenario, rendering the basis for the analysis and design of MIMO systems. Due to the particular properties of LIS, the model accounts for superdirectivity and mutual coupling effects along with near field propagation, necessary in those situations where the array dimension becomes very large. Finally, with the proposed model, the matched filter transmitter and the weighted  minimum  mean  square  error precoding are derived under both realistic constraints: limited radiated power and maximum ohmic losses.  
\end{abstract}

%% file: introduction.tex
Since the seminal paper by Marzetta \cite{Marzetta2010}, massive \ac{MIMO} systems have moved from being an unrealistic idea to becoming a key enabling technology in 5G and future generations of wireless networks \cite{Agiwal2016, Bjornson2019}. The promising gain of these systems have given raise to a widespread interest in considering even a larger number of antennas than in conventional massive \ac{MIMO}. Hence, new concepts such as \textit{holographic \ac{MIMO}}, \textit{\ac{LIS}} or \textit{\ac{IRS}} have emerged as a natural evolution of classical \ac{MIMO}.

The use of \ac{LIS} (i.e., large arrays) for wireless networks may render considerable gains in terms of capacity, interference reduction and user multiplexing; but it also supposes a new paradigm from a system design point of view. Introducing a massive number of antennas in a limited surface leads to a small inter-element distance (ideally almost-continuous radiating surfaces \cite{dardari_communicating_2019}). Hence, phenomena that have been classically neglected in the analysis of \ac{MIMO} systems, such as \textit{mutual coupling} \cite{Morris2005, Harrington1965, Lo1966} and \textit{superdirectivity effect} \cite{Schelkunoff1943, uzsoky_theory_1956, Morris2005, abdallah_maximum_2016, marzetta_super-directive_2019}, become now much relevant. 

Mutual coupling is inherent to arrays with closely-spaced antennas, affecting both the radiation pattern and the impedance of the antenna element, which implies ultimately a change on the received power \cite{balanis}. However, this effect is not considered when designing the linear transmit and receive processing \cite{ christensen_weighted_2008, Joham2005}
, giving rise to solutions that might not be optimal in realistic conditions. Also, related to mutual coupling is the superdirectivity effect, which theoretically allows for the design of highly directive (ideally unbounded) arrays of closely-spaced antennas \cite{uzsoky_theory_1956}. However, in practice, achieving such superdirectivity comes at the price of extremely large excitation currents, which considerably increases the losses and reduces the efficiency \cite{Schelkunoff1943}, and makes the array sensitive to small random variations in the excitation \cite{gilbert_optimum_1955}.

On a related note, as the array dimensions become large and the number of the antennas increases, some of the classical results for \ac{MIMO} systems are no longer valid. For instance, in \cite{Bjornson2020}, it is proved that the widely accepted scaling law, i.e., the signal-to-noise ratio scales with the number of antennas, is only valid under the far-field assumption. Therefore, in order to properly analyze and design \ac{LIS}-based \ac{MIMO} systems, it is also necessary to consider \textit{near-field} effects, specially in indoor scenarios or those situations where far-field conditions cannot be guaranteed due to the large \ac{LIS} dimensions.

Although some works have considered the effects of superdirectivity arrays \cite{Morris2005, Bikhazi2007} or mutual coupling \cite{Wallace2004}, to the best of our knowledge, no model accounting for superdirectivity, coupling and near-field propagation has been presented in the literature. Aiming to fill this gap, we here propose a communication model for \ac{LIS}-based \ac{MIMO}, which considers the three aforementioned phenomena. To that end, we merge electromagnetic theory with classical \ac{MIMO} system models, creating a link that allows to include all these effects in the channel matrix and paving the way to more detailed works. As a result, we obtain a characterization based on infinitesimal dipoles, which is independent of any physical antenna realization and can be used to model real deployments, e.g., metasurfaces \cite{Shlezinger2019}. Finally, we use the derived model to explore the design and performance of two transmission schemes: \ac{MF} and \ac{WMMSE} \cite{christensen_weighted_2008}. 

\textit{Notation:} $i$ is the imaginary unit, $\Vert\cdot\Vert_2$ is the euclidean norm, $\vert\cdot\vert$ is the absolute value and $\cdot^T$ and $\cdot^H$ are the transpose and Hermitian transpose respectively. Vectors are denoted by bold lowercase symbols, and matrices are denoted by bold uppercase symbols. Finally, $\Re\{\cdot\}$,  $\Tr{\cdot}$ and $\mathbb{E}[\cdot]$ are the real part, the trace and the expectation operator, respectively.

%% file: systemModel.tex
We consider a downlink multi-user \ac{MIMO} system in which a base station communicates with $M$ \acp{UE}. All the users are equipped with single-antenna devices, whilst an \ac{LIS} is deployed at the base station. The \ac{LIS} is modelled as a collection of $N$ closely spaced antennas, emulating a near-continuous radiating surface, and centered at the origin of a cartesian coordinate system aligned with the $yz$-plane, whereas the \acp{UE} are arbitrary placed in front of it. 

The antennas composing the \ac{LIS} are modelled as identical and infinitesimal dipoles carrying a uniform current along a short line segment, where, by definition, the current distribution is independent of the surroundings. Note that, with this model, we are abstracting from any physical structure for the antennas, and just considering the antennas as uniform current sources where the voltage is simply the difference in electrical potential along the length of the dipoles. 
This keeps the mathematical complexity of the model under control and allows to capture near-field propagation effects without resorting to complicated electromagnetic simulations. Also, we only consider linear $z$-polarized receivers and transmitters, and the effects of the near-field cross-polarization terms is left for future work. As in \cite{dardari_communicating_2019, Bjornson2020}, we assume a pure \ac{LoS} propagation scenario, neglecting fading and shadowing.  

\begin{figure}
    \centering
    \vspace{0.03in}
    \includegraphics[scale = 1]{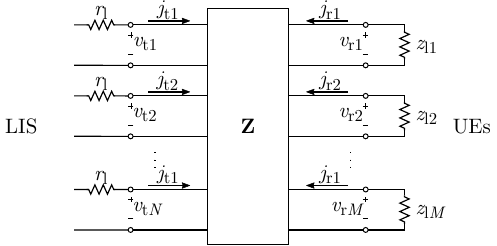}
    \caption{Circuit model of the scenario as a multi-port network. The ports on the left represent antennas in the \ac{LIS} where the currents $j_{\text{t}n}$ run through the loss resistors $r_\text{l}$ before entering the network. The ports on the right represent the \acp{UE} which are terminated in load impedances $z_{\text{l}m}$.}
    \label{fig:network}
\end{figure}

To address the impact of near-field propagation and superdirectivity effects, we consider a circuital model for the aforementioned \ac{MIMO}  system, similarly as done in \cite{Wallace2004} to analyze mutual coupling. In the model, represented in Fig. \ref{fig:network}, every antenna element in the \ac{LIS} and every \ac{UE} is represented by individual ports carrying different currents and voltages. To model ohmic losses within the \ac{LIS}, of capital importance in superdirective systems \cite{Schelkunoff1943}, identical loss resistances $r_\text{l}$ are attached to every \ac{LIS} port. On the receiver side, all \ac{UE} ports are terminated in load impedances $z_{\text{l}m}$. The relation between the currents and voltages is therefore given by
\begin{align}
    \begin{bmatrix}\vec{v}_\text{t} \\ \vec{v}_\text{r} \end{bmatrix} = \underbrace{\begin{bmatrix}\mat{z}_\text{tt} & \mat{z}_\text{rt}^T \\ \mat{z}_\text{rt} & \mat{z}_\text{rr} \end{bmatrix}}_{\mat{z}}  \begin{bmatrix}\vec{j}_\text{t} \\ \vec{j}_\text{r} \end{bmatrix}, \label{eq:system}
\end{align}
where $\vec{j}_\text{t}\in\mathbb{C}^{N\times 1}$ and  $\vec{j}_\text{r}\in\mathbb{C}^{M\times 1}$ are the currents vectors in the \ac{LIS} antenna elements and the \acp{UE},  $\vec{v}_\text{t}\in\mathbb{C}^{M\times 1}$ and $\vec{v}_\text{r}\in\mathbb{C}^{M\times 1}$ are the voltage vectors across the \ac{LIS} and \acp{UE} ports, and $\mat{Z}\in\mathbb{C}^{(N+M)\times (N+M)}$ is the system impedance matrix, which can be split into different submatrices. Specifically, $\mat{Z}_\text{tt}\in\mathbb{C}^{N\times N}$ is the \ac{LIS} impedance matrix representing the mutual coupling between the different antenna elements, $\mat{z}_\text{rr}\in\mathbb{C}^{M\times M}$ represents the coupling between the \acp{UE}, and $\mat{z}_\text{rt}\in\mathbb{C}^{M\times N}$ is the \ac{LIS} to \ac{UE} impedance matrix, capturing the propagation effects. Eq. \eqref{eq:system} is the basis of this paper, allowing us to create a link between the electromagnetic theory and the discrete models widely used in communications. 

%% file: systemAnalysis.tex
\subsection{Transmitted power, received power and efficiency} 

From the circuital model in Fig. \ref{fig:network}, the signal power at the receivers is equal to the power dissipated in the attached loads $z_{\text{l}m}$ ($m=1,\dots,M$). By Eq. \eqref{eq:system}, the voltage across the \ac{UE} ports is given as the sum of the \ac{LoS} propagation and the scattered waves originating from the \acp{UE}, i.e.,
\begin{align}
    \vec{v}_\text{r} = \underbrace{\mat{Z}_\text{rt} \vec{j}_\text{t}}_{\text{LoS}} + \underbrace{\mat{Z}_\text{rr} \vec{j}_\text{r}}_{\text{scattering}}. \label{eq:receivedpower1}
\end{align}

Also, applying Ohm's law at the receiver ports, the received voltage is expressed as
\begin{align}
    \vec{v}_\text{r} = -\mat{z}_\text{l}\vec{j}_\text{r}, \label{eq:Ohmlawreceiver}
\end{align}
where $\mat{z}_\text{l}\in\mathbb{C}^{M \times M}$ is a diagonal matrix with the $m$-th diagonal element equal to the load impedance $z_{\text{l}m}$. We consider that the \acp{UE} are spaced such that the impedance looking into the multiport network is dominated by the antenna's self-impedance $z_0$ and, therefore, we perform a conjugate matching of the self-impedance, i.e.,  $z_{\text{l}m}=z_0^*$ $\forall$ $m$. Introducing \eqref{eq:Ohmlawreceiver} in \eqref{eq:receivedpower1}, the relation between the transmitted and received currents is  given by 
\begin{align}
    \vec{j}_\text{r}= -(\mat{Z}_\text{rr} + \mat{I}_{M} z_0^*)^{-1}\mat{Z}_\text{rt} \vec{j}_\text{t}. \label{eq:recCurrents}
\end{align}
With the relation between $\vec{j}_\text{r}$ and $\vec{j}_\text{t}$, the time-averaged power received at the $m$-th \ac{UE} is directly expressed as
\begin{align}
    P_{\text{r}m} = \frac{\Re{-{j}_{\text{r}m}^H {v}_{\text{r}m}}}{2}=\frac{\left|j_{\text{r}m}\right|^2 \Re{z_{0}}}{2}, \label{eq:powerRecieved}
\end{align}
where $j_{\text{r}m}$ and $v_{\text{r}m}$ for $m=1,\dots,M$ are the elements of $\vec{j}_{\text{r}}$ and $\vec{v}_{\text{r}}$, respectively.  

On the transmitter side, the primary interest is the time-averaged power delivered to the network, which is given by\footnote{As the impedance matrices are symmetric and have equal diagonal elements, the currents can isolated outside the real part operator.}
\begin{align}
   P_\text{t} &= \underbrace{\frac{\Re{\vec{j}_\text{t}^H \mat{Z}_\text{tt} \vec{j}_\text{t}}}{2}}_{\text{internal}} + \underbrace{\frac{\Re{\vec{j}_\text{t}^H \mat{Z}_\text{rt}^T \vec{j}_\text{r}}}{2}}_{\text{external}} =\frac{\vec{j}_\text{t}^H \mat{R}_\text{P} \vec{j}_\text{t}}{2} \label{eq:powerTransmitted}, \\
   \mat{R}_\text{P} &= \Re{\mat{Z}_\text{tt}-\mat{Z}_\text{rt}^T\left(\mat{Z}_\text{rr}+\mat{i}_M z_0^*\right)^{-1} \mat{Z}_\text{rt}}.
\end{align}

In \eqref{eq:powerTransmitted}, the first term (labeled as internal) is the actual power that the transmitter delivers to the network, which is only impacted by the coupling matrix between the antenna elements in the \ac{LIS}. The second term encompasses the coupling between the inducted currents at the \acp{UE} and the \ac{LIS}, which may be relevant in near-field scenarios where the distance between the users and the transmitter is small. If all the \acp{UE} are placed far enough, then this second term can be neglected. 

Finally, to model thermal losses at the transmitter, a loss resistance $r_\text{l}$ is attached to all ports corresponding to the \ac{LIS} antennas. These losses, although usually ignored in \ac{MIMO} works, play a pivotal role in beamforming analysis and design. As the antennas in the \ac{LIS} are very closely spaced, optimal precoders result in very high currents, which leads to significant thermal losses even for very high efficiency antennas \cite{Schelkunoff1943, Harrington1965, gilbert_optimum_1955}. With known currents $\vec{j}_\text{t}$, the thermal losses are given by
\begin{align}
    P_l = \frac{\vec{j}_\text{t}^H r_\text{l} \mat{I}_{M} \vec{j}_\text{t}}{2} = \frac{r_\text{l} \vec{j}_\text{t}^H \vec{j}_\text{t}}{2} \label{eq:thermalLoss}
\end{align}
and, therefore, the radiation efficiency of an isolated antenna can be expressed in terms of its self-impedance and loss impedance as
\begin{align}
    e_r = \frac{\Re{z_0}}{\Re{z_0}+r_\text{l}}. \label{eq:radEffSingle}
\end{align}

\subsection{Antenna modelling and inter-element coupling} 
\label{sec:EM}

All the results in the previous subsection are in terms of the impedance matrices of the system. The entries of these matrices depends on the specific physical realisation of the antennas, their distance, and their orientation. As stated before, in this work the antennas are modelled as small dipoles carrying uniform currents. To derive the coupling between them (i.e., the impedance matrix elements), a single radiating antenna positioned at the origin is modelled as a source field $\mat{J}$. The radiated electrical field is then given in terms of the Green's tensor function \cite[Eq. (3-65)]{harrington_time-harmonic_2001} as 
\begin{align}
    \mat{E}(\vec{r})&=\iiint_V \mat{G}(\vec{r}-\vec{r}') \mat{J}(\vec{r'}) \dif \vec{r'}, \label{eq:Eint}\\
    \mat{G}(\vec{r}) &= -i\frac{\eta}{2\lambda}\left(\mat{I}_3 + \frac{1}{k^2} \nabla\nabla^T\right) \frac{\e{-i k r}}{r}, \label{eq:Gtensor}
\end{align}
where $\eta$ is the free space impedance, $\lambda$ is the wavelength, $k = 2\pi/\lambda$ denotes the wavenumber, $\vec{r} = \begin{bmatrix} x & y & z\end{bmatrix}^T$ is a distance vector defined by its cartesian coordinates with norm $r = \|\vec{r}\|_2$, and $\nabla = \begin{bmatrix}\pdv{x} & \pdv{y} & \pdv{z}\end{bmatrix}^T$ is the gradient operator. 
Since each antenna is a line source of length $l$ carrying a uniform current $j$ along the $z$-direction, the radiated field expression reduces to 
\begin{align}
    \mat{E}(\vec{r})&=\int_{-\frac{l}{2}}^{\frac{l}{2}} \mat{G}\left(\vec{r}- \begin{bmatrix}0&0&z'\end{bmatrix}^T\right) \begin{bmatrix}0&0&j\end{bmatrix}^T \dif z',\label{eq:fieldintegral}
\end{align}
which, by assuming the length of the dipole is short, is approximated as
\begin{align}
    \mat{E}(\vec{r}) \approx \mat{G}\left(\vec{r}\right) \begin{bmatrix}0&0&j l\end{bmatrix}^T. \label{eq:eFieldSum}
\end{align}

At a receiver, the voltage across the receiving antenna is given by integration along the line segment of the receiver as
\begin{align}
    v(\vec{r}) = 
    -\int_{\frac{-l}{2}}^{\frac{l}{2}} \begin{bmatrix}0&0&1\end{bmatrix} \mat{E}(\vec{r}) \dif z, \label{eq:indVoltage}
\end{align}
which, assuming again a short dipole, can be approximated as
\begin{align}
     v\left(\vec{r}\right) \approx - \begin{bmatrix}0&0&l\end{bmatrix} \mat{G}\left(\vec{r}\right) \begin{bmatrix}0&0&j l\end{bmatrix}^T. \label{eq:indVoltage2}
\end{align}

Finally, dividing \eqref{eq:indVoltage2} by the source current yields the mutual impedance between two antennas separated by the distance vector $\vec{r}$ as
\begin{align}
    z(\vec{r}) =i\frac{l^2\eta \e{-ikr}}{2\lambda r} \left( 1 - \frac{z^2}{r^2} - \frac{i}{k r} - \frac{1}{k^2 r^2} + \frac{i 3 z^2}{k r^3}  + \frac{3 z^2}{k^2 r^4} \right). \label{eq:impfunction}
\end{align}

In our proposal, all the \ac{UE} and \ac{LIS} antennas are modelled as small dipoles and, hence, the entries of the impedance matrix $\mat{z}$ in  \eqref{eq:system} are given by \eqref{eq:impfunction}. Similarly, introducing \eqref{eq:impfunction} in \eqref{eq:powerRecieved}, we obtain the received power by an arbitrary user.

A consequence of the infinitesimal antenna model is that the imaginary part of the antenna's self-impedance diverges to $\pm\infty$ depending on the direction of approach. This means that it is practically impossible to perform an impedance matching for an infinitesimal dipole. However, the model can be seen as a discretization of a physical system which could potentially be realised. As such, the model serves to represent an arbitrary design that is theoretically possible while being independent of practical implementation  limitations. For instance, the same technique is observed in \cite{mikki_theory_2007}, where  common antenna designs are split into sets of infinitesimal dipoles while successfully capturing the narrow-band radiation characteristics of the original design.

As shown in \eqref{eq:powerRecieved} and \eqref{eq:powerTransmitted}, the system performance is dominated by the real part of the impedance, whose maximum value is obtained as $\vec{r}\rightarrow\vec{0}$ (corresponding to the value of $z_0$), i.e., 
\begin{align}
    \Re{z_0} = \lim_{\vec{r} \to \vec{0}} \left(  \Re{z\left(\vec{r}\right)}  \right) = \frac{k l^2 \eta}{3 \lambda}. \label{eq:radResistance}
\end{align}
Without any loss of generality, the length of the short dipoles is chosen as to normalize the radiation resistance, and thus $l = \sqrt{\frac{3\lambda}{k \eta}} \approx 0.036 \lambda$.

%% file: MIMOcommunmodel.tex
With the analysis of the circuital model in Fig. \ref{fig:network} accomplished, the next step is linking it to a \ac{MIMO} communication model that can be easily used for precoding analysis and design. To that end, we consider that the base station serves $M$ users simultaneously, and therefore the decoded signal vector $\widehat{\vec{x}} \in\mathbb{C}^{M\times 1}$ is expressed as 
\begin{align}
    \widehat{\vec{x}} = \mat{a} \left( \mat{h}\mat{b}\vec{x} + \vec{n} \right), \label{eq:MIMOmodel}
\end{align}
where each element is explained in the following. Vector $\vec{x} = \begin{bmatrix}x_1 & x_2 & \dots & x_M\end{bmatrix}^T$ denotes the set of symbols intended to each user, represented as complex \ac{RMS} values with zero mean and covariance matrix  $\mathbb{E}\left[ \vec{x}\, \vec{x}^H \right]=\mat{I}_{M}$, where $x_m \in \mathbb{C}$ denotes the symbol destined for $m$-th user. This set of symbols is passed through a transmit filter (precoding) represented by matrix $\mat{B} = \begin{bmatrix}\vec{b}_1 & \dots & \vec{b}_M\end{bmatrix}$, with $\vec{b}_m\in\mathbb{C}^{N\times 1}$ the beam targeted at $m$-th user. Coming back to the circuital model in \eqref{eq:system}, the \ac{RMS} value of the currents at the transmitter would be given then by $\vec{j}_{\text{t}}/\sqrt{2} = \mat{B}\vec{x}$. Note that these currents are time varying, but for simplicity we remove the temporal dependency. Given $\vec{j}_\text{t}$, the currents induced at the receivers are given by \eqref{eq:recCurrents}, and, therefore, the channel matrix is expressed as 
\begin{equation}
    \mat{H} = -(\mat{Z}_\text{rr} + \mat{I}_{M} z_0^*)^{-1}\mat{Z}_\text{rt} = \begin{bmatrix}\vec{h}_1 & \vec{h}_2 & \dots & \vec{h}_M \end{bmatrix}^H, \label{eq:channel}
\end{equation}
where $\vec{h}_m\in\mathbb{C}^{N\times 1}$ is the channel vector from the \ac{LIS} to $m$-th user. At the receiver side, a diagonal filter matrix $\mat{A}\in\mathbb{C}^{M\times M}$ is applied to the received symbols. Finally, $\vec{n}\sim\mathcal{CN}_M(\mat{0}_{M\times 1}, \sigma_n^2\mat{I}_M)$ is the noise term, which is independent of the transmitted symbols, i.e., $\mathbb{E}[\vec{n}\vec{x}^H] = \mat{0}_{M\times M}$.   

Note that we have presented a formulation in terms of received and transmitted currents, but analogous formulations in terms of voltages can be obtained by applying the relations between them, e.g., \eqref{eq:Ohmlawreceiver}, giving raise to models as in \cite{Morris2005}. 

Introducing this communication model in \eqref{eq:powerTransmitted} and \eqref{eq:thermalLoss}, the radiated power and the thermal losses are thus rewritten as 
\begin{align}
    P_\text{t} =&\Tr{ \text{E}\left[ \vec{x}^H \mat{b}^H \mat{R}_\text{P} \mat{b} \vec{x}  \right] } =\Tr{\mat{b}^H \mat{R}_\text{P} \mat{b}}, \label{eq:multiPtx}\\
    P_l =& r_l\Tr{ \text{E}\left[ \vec{x}^H \mat{b}^H \mat{b} \vec{x}  \right] } = \frac{1 - e_r}{e_r}\Tr{\mat{b}^H \mat{b}}. \label{eq:multiPl}
\end{align}

Moreover, two of the key metrics for beamforming design, namely the expectation of the \ac{SINR} of the $m$-th user and the maximal achievable sum rate, are given as
\begin{align}
    \rho_m &=\frac{ \left| \vec{H}_{m}^H \vec{B}_m \right|^2 }{\sum_{n\neq m}^M \left|\vec{H}_m^H \vec{B}_n\right|^2  + \sigma_n^2}, \label{eq:sinr} \\ 
    C &=\sum_{m = 1}^M \log_2\left(1 + \rho_m\right). \label{eq:sumCap}
\end{align}

%% file: transmitDesign.tex
Based on the communication model in \eqref{eq:MIMOmodel}, we here explore the transmit design through two possible implementations: \ac{MF} and the \ac{WMMSE} defined in \cite{christensen_weighted_2008}. For both transmitters, we back off from the widely-used constraint of limited power, and consider two constraints that are of interest when designing \ac{LIS}-based communications: \textit{radiated power} \eqref{eq:multiPtx} and \textit{ohmic losses} \eqref{eq:multiPl}. Considering the radiated power constraint instead of the traditional $\Tr{\mat{b}^H \mat{b}}$ is important in highly coupled systems, since the latter may lead to a considerably larger actual radiated power \cite{Morris2005}. On the other hand, considering ohmic losses is necessary since these losses are usually high in superdirective systems \cite{Lo1966}, and it seems more realistic and feasible than restraining the superdirectivity $Q$ factor \cite{Bikhazi2007}.

\subsection{\ac{MF} transmitter}

For the matched filter, we follow the objective of \cite{Joham2005} in maximizing the correlation between the received and transmitted symbol. The diagonal receive filter $\mat{A}$ is set to the identity matrix $\mat{I}_m$ as it does not affect the correlation. The optimization problem is then formulated as
\begin{maxi}[2]
{\mat{B}}{\text{E}\left[ \vec{x}^H \widehat{\vec{x}} \right] = \Tr{\mat{H}\mat{B}}}{\label{eq:MF}}{}
\addConstraint{\Tr{\mat{B}^H \mat{R}_\text{P} \,\mat{B}}} {\leq P_R}
\addConstraint{r_\text{l} \Tr{\mat{B}^H \,\mat{B}}} {\leq P_L},
\end{maxi}
where $P_R$ and $P_L$ are the maximum allowed radiated power and ohmic losses, respectively.
The \ac{KKT} conditions are given as
\begin{align}
    \mat{B} &= \left(2\mu_1 \mat{R}_\text{P} + 2\mu_2 r_\text{l} \mat{i}_N\right)^{-1}\mat{H}^H, \label{eq:lagrangianDerivative}\\
    P_R&=\Tr{\mat{B}^H \mat{R}_\text{P} \mat{B}}, \label{eq:MFKKT2}\\
    P_L&= r_\text{l}\Tr{\mat{B}^H \mat{B}}, \label{eq:MFKKT3}
\end{align}
where \eqref{eq:lagrangianDerivative} is obtained by setting the derivative of the Lagrangian function with respect to $\mat{b}$ to zero and isolating for $\mat{B}$, and $\mu_1>0$ and $\mu_2>0$ are the Lagrangian multipliers. The three possible solutions are given as follows. The first two options where either $\mu_1 = 0$ or $\mu_2 = 0$ yield the thermal loss constrained and radiated power constrained solutions as
\begin{align}
\evalat*{\mat{B}_\text{MF}}{\mu_1=0}&=\mat{H}^H \sqrt{\frac{P_L}{\Tr{\mat{h}\,\mat{h}^H r_\text{l}}}}, \label{eq:MFLossConstrained}\\
\evalat*{\mat{B}_\text{MF}}{\mu_2=0}&=\mat{R}_\text{P}^{-1}\mat{H}^H \sqrt{\frac{P_R}{\Tr{\mat{h}\,\mat{R}_\text{P}^{-1}\mat{h}^H}}}. \label{eq:MFRadConstrained}
\end{align}

To the best of authors' knowledge, if both $\mu_1 \neq0$ and $\mu_2 \neq 0$, a closed form solution cannot be obtained. Instead, we propose an algorithm which rapidly converges to a solution within a specified precision. To this end, a variable $\alpha = \frac{\mu_2}{\mu_1}$ is defined and inserted into \eqref{eq:lagrangianDerivative}, leading to
\begin{align}
    \evalat*{\mat{B}_\text{MF}}{\mu_1, \mu_2 \neq 0} &= \left(\mat{R}_\text{P} +\alpha r_\text{l} \mat{i}_N\right)^{-1} \mat{H}^H\frac{1}{2\mu_1}. \label{eq:MFalpha}
\end{align}

Using \eqref{eq:MFalpha}, \eqref{eq:MFKKT2} and \eqref{eq:MFKKT3} are rewritten as
\begin{align}
    4 P_R \mu_1^2&=\Tr{\mat{H}\left(\mat{R}_\text{P} +\alpha r_\text{l} \mat{i}_N\right)^{-1} \mat{R}_\text{P} \left(\mat{R}_\text{P} +\alpha r_\text{l} \mat{i}_N\right)^{-1} \mat{H}^H}, \label{eq:MF4}\\
    4 P_L \mu_1^2&= r_\text{l}\Tr{\mat{H}\left(\mat{R}_\text{P} +\alpha r_\text{l} \mat{i}_N\right)^{-2} \mat{H}^H}. \label{eq:MF5}
\end{align}
The optimal value of $\alpha$ is obtained by dividing \eqref{eq:MF4} and \eqref{eq:MF5}, 
\begin{align}
    \frac{P_L}{P_R} = \frac{r_\text{l}\Tr{\mat{H}\left(\mat{R}_\text{P} +\alpha r_\text{l} \mat{i}_N\right)^{-1} \left(\mat{R}_\text{P} +\alpha r_\text{l}\mat{i}_N\right)^{-1} \mat{H}^H}}{\Tr{\mat{H}\left(\mat{R}_\text{P} +\alpha r_\text{l} \mat{i}_N\right)^{-1} \mat{R}_\text{P} \left(\mat{R}_\text{P} +\alpha r_\text{l} \mat{i}_N\right)^{-1} \mat{H}^H}} \label{eq:MF6}
\end{align}
and solving numerically. As the right hand side of \eqref{eq:MF6} is a monotonically decreasing function in the variable $\alpha$, it is easily solved using numerical methods. Once $\alpha$ is obtained, $\mu_1$ can be determined from either \eqref{eq:MF4} or \eqref{eq:MF5}. Finally, the \ac{MF} beamforming matrix is calculated as in \eqref{eq:MFalpha}.

\subsection{\ac{WMMSE} transmitter}

The \ac{MMSE} transmitter was derived for a single quadratic constraint in \cite{Joham2005}. Later, the authors in \cite{christensen_weighted_2008} showed that maximizing the weighted sum rate is equivalent, from an optimization point of view, to the \ac{WMMSE} problem when the weights are chosen in an optimal way. They also proposed an iterative algorithm for jointly optimizing the \ac{MMSE} beamforming matrix and receive weights. We here follow a procedure similar to that in \cite{christensen_weighted_2008} to derive the \ac{WMMSE}, but including both constraints (radiated power and ohmic losses), as done in the \ac{MF} transmitter,  and introducing some elements from \cite{Joham2005}. Hence, the \ac{WMMSE} optimization problem is formulated as
\begin{mini}[2]
{\mat{A}, \mat{B}, \mat{W}, \beta}{\text{E}\left[ \left\Vert \mat{w}^{\frac{1}{2}} \left(\vec{x} - \frac{1}{\beta} \widehat{\vec{x}}\right)\right\Vert_2^2\right],}{\label{eq:MMSE}}{}
\addConstraint{\Tr{\mat{B}^H \mat{R}_\text{P} \,\mat{B}}} {\leq P_R}
\addConstraint{r_\text{l} \Tr{\mat{B}^H \,\mat{B}}} {\leq P_L},
\end{mini}
where $\mat{W}\in\mathbb{R}^{M\times M}$ is a diagonal weighting matrix, and $\beta$ is a positive constant. As in  \cite{Joham2005}, $\beta$ is used to enforce the powers constraints. To that end, we set the derivative of the Lagrangian function w.r.t. $\mat{B}$ to zero, obtaining
\begin{equation}
     \mat{B} = \beta\underbrace{\left( \mat{H}^H \mat{A}^H \mat{W} \mat{A} \mat{H} + \alpha_1 \mat{R}_\text{P} + \alpha_2 r_\text{l} \mat{I}_N   \right)^{-1} \mat{H}^H \mat{A}^H \mat{W}^H}_{\makebox[0pt]{\text{$\widetilde{\mat{B}}$}}}, \label{eq:optiB}
\end{equation}
with $\alpha_{1,2} = \beta^2\mu{1,2}$, where $\mu{1,2}$ are the Lagrange multipliers for both constraints in \eqref{eq:MMSE}. Then, introducing \eqref{eq:optiB} in the power constraints equations yields
\begin{align}
\beta_L = \sqrt{\frac{P_L}{\Tr{r_\text{l}\widetilde{\mat{B}}^H \widetilde{\mat{B}}}}}  ~~\text{and}~~  \beta_{R} = \sqrt{\frac{P_R}{\Tr{\widetilde{\mat{B}}^H \mat{R}_\text{P} \widetilde{\mat{B}}}}}, 
\end{align}
and the value of $\beta$ is chosen so that both constraints are satisfied, i.e., $\beta = \min\{\beta_L,\beta_R\}$. 

Regarding the values of $\alpha_1$ and $\alpha_2$, the corresponding value for a single power constraint is calculated in  \cite{Joham2005} as $\alpha = \sigma_n^2 \Tr{\mat{A}^H \mat{W} \mat{A}}/P$, where $P$ is the power constraint. Based on this result, we propose an heuristic solution given by 
\begin{align}
    \alpha_1 = \frac{\sigma_n^2 \Tr{\mat{A}^H \mat{W} \mat{A}}}{P_R}  ~~\text{and}~~  \alpha_2 = \frac{\sigma_n^2 \Tr{\mat{A}^H \mat{W} \mat{A}}}{P_L}. \label{eq:alphas}
\end{align}
Finally, conditioned on $\mat{B}$, the elements of the diagonal matrices $\mat{A}$ and $\mat{W}$, namely $a_m$ and $w_m$ for $m=1,\dots,M$, are obtained as \cite[eqs. (7) and (38)]{christensen_weighted_2008}
\begin{align}
    a_m &=\vec{B}_{m}^H\vec{H}_{m} \left( \vec{h}_{m}^H\vec{b}_{m}\vec{b}_{m}^H\vec{h}_{m} + r_m \right)^{-1},\label{eq:Adiagonal} \\
    w_m &= 1 +  \vec{b}_{m}^H\vec{h}_{m} r_m^{-1} \vec{h}_{m}^H \vec{b}_{m}, \label{eq:weights}
\end{align}
with 
\begin{equation}
    r_m = 1 + \sum_{i=1, i\neq m}^M\left( \vec{H}_{m}^H \vec{B}_{i} \vec{B}_{i}^H \vec{H}_{m} \right). \label{eq:rcovariance}
\end{equation}

With all the variables involved in \eqref{eq:MMSE} defined, the proposed iterative procedure to compute the \ac{WMMSE} transmitter is summarized in Algorithm \ref{alg:WMMSE}. 
\RestyleAlgo{ruled}
\begin{algorithm}
  \kwInit{$\mat{a}^0 = \mat{w}^0 = \mat{I}_M$, $n=0$}
   \Repeat{convergence}{
   I. Calculate $\mat{B}^n$ from \eqref{eq:optiB}-\eqref{eq:alphas}\\
   II. Update diagonal elements of $\mat{A}^{n+1}$ using \eqref{eq:Adiagonal} and \eqref{eq:rcovariance}\\
   III. Update diagonal elements of $\mat{W}^{n+1}$ using \eqref{eq:weights}\\
   IV. Update $n=n+1$\\}
  \caption{WMMSE}
  \label{alg:WMMSE}
\end{algorithm}

%% file: simulation.tex
Finally, we present some simulated results for both proposed beamformers, namely \ac{MF} and \ac{WMMSE}, in order to show the impact of coupling, antenna efficiency and number of radiating elements in the number of supported simultaneous \acp{UE}. Throughout the whole section, perfect knowledge of the channel in \eqref{eq:channel} is assumed. 

\begin{figure}
    \centering
    \includegraphics[scale=0.86]{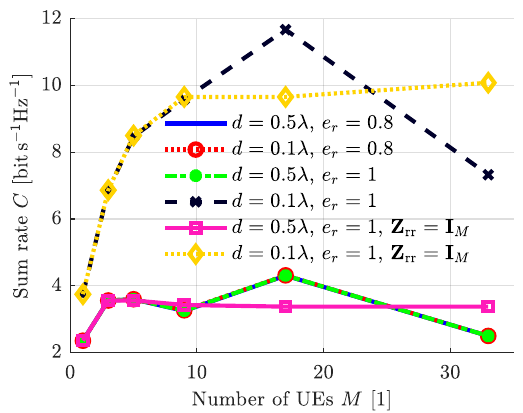}
        \vspace{-10pt}
    \caption{Simulated \ac{WMMSE} sum capacity for a linear one-dimensional \ac{LIS} and a linear formation of \acp{UE}.  The noise variance is set to $\sigma_n = 10^{-4}$. The minimum inter-\ac{UE} spacing ($M =33$) is $0.3125\lambda$. The radiated power and loss power constraints are set to $P_R=1$ and $P_L=1$.}
    \label{fig:WMMSEresults}
\end{figure}

We consider first a one-dimensional transmit array with length of $4\lambda$ along the $y$-axis, populated by either $N=9$ or $N=41$ transmitters, corresponding to a spacing of $d=0.5\lambda$ and $d=0.1\lambda$, respectively. The \acp{UE} are positioned at a distance of $20\lambda$ along the $x$-axis on a line along the $y$-axis of length $10\lambda$. Hence, the larger the number of \acp{UE}, the smaller the separation between them.
In this scenario, the \ac{WMMSE} sum capacity in \eqref{eq:sumCap} is calculated in terms of the number of users with difference antenna efficiency and spacing between the \ac{LIS} elements, as depicted in Fig. \ref{fig:WMMSEresults}. We observe that the inter-\ac{UE} coupling plays an important role in the performance. If we neglect it, i.e.  $\mat{Z}_\text{tt} = \mat{i}_M$, then the sum capacity rises to a maximum and remains constant as more users are added. However, if we take into account this coupling, then the capacity raises up to a turning point, from which it starts decreasing as the number of users increases. Note, however, that the maximum value for the capacity is higher when coupling is present.  Notably, with a relatively low antenna efficiency of $e_r = 0.8$, decreasing the spacing from $d = 0.5\lambda$ to $d=0.1\lambda$ yields no gains in terms of sum capacity since the system is limited by ohmic losses. 

\begin{figure}
\vspace{-10pt}
    \centering
    \includegraphics[scale=0.86]{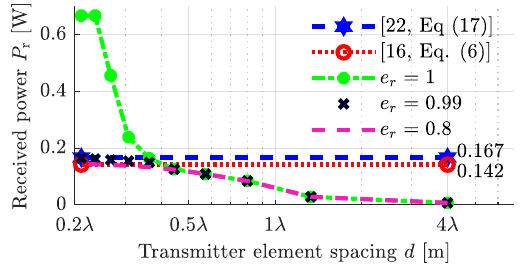}
    \vspace{-10pt}
    \caption{Simulated \ac{MF} received power at a \ac{UE} positioned two wavelengths from a four-by-four wavelength \ac{LIS}. The radiated power and loss power constraints are set to $P_R=1$ and $P_L=1$.}
    \label{fig:MFResults}
    \vspace{-10pt}
\end{figure}

On the other hand, Fig. \ref{fig:MFResults} shows the received power from a four-by-four wavelength two-dimensional transmit array, populated by an increasing number of transmitters. A \ac{UE} is aligned within the center of the \ac{LIS} at a distance of $2\lambda$ along the $x$-axis. The received power at the \ac{UE} is determined using the \ac{MF} transmitter for three different efficiencies, whilst the transmit and loss power constraints are fixed to $P_R = 1$ and $P_L = 1$. As a reference, we represent also the results obtained from the models in \cite{Hu2018} and \cite{Bjornson2020}.  In this scenario, the \ac{LIS} covers one sixth of all space seen from the perspective of the \ac{UE}. As such, the model of \cite{Hu2018}, which models the physical aperture of the \ac{LIS} as an electrical aperture, yields a received power of ${P_R}/{6}$. Regarding Fig. \ref{fig:MFResults}, we see that, when $e_r < 1$, the performance approaches the result predicted by \cite{Hu2018} as the number of antenna-elements increases (equivalently, $d$ decreases). In turn, for $e_r = 1$, the super-directivity effect is not restricted, allowing electrical aperture to extend beyond the physical aperture and reach a point where the scattering from the \ac{UE}, marked as external in \eqref{eq:powerTransmitted}, limits the radiated wave. Note that, if the scattering is not included in the model, the received power can extend beyond the transmitted power and energy is not conserved.

%% file: conclusion.tex
We have introduced a new communication model for multi-user \ac{MIMO} based on \ac{LIS}, which takes into account several phenomena that has been classically neglected in \ac{MIMO} analysis and design such as mutual coupling, superdirectivity and near-field effects. The proposed model, which merge communication and electromagnetic theory, may be a first step in the design of \ac{LIS} systems in realistic conditions. Based on our proposal, we also provide, for first time in the literature, a transmit \ac{MF} and \ac{WMMSE} beamforming schemes that take into account two constraints: effective radiated power and ohmic losses. As shown in the provided numerical results, the impact of near-field propagation (namely, coupling between the users and the transmitter) as well as that of the antenna efficiency and ohmic losses are non-negligible in the system performance, so its characterization seems to be of key importance in the future design of \ac{LIS}-based solutions.